\documentclass[a4paper]{jpconf}

\usepackage{citesort}

\usepackage{graphicx}
\begin{document}
\title{Quantum-classical interactions and measurement: a consistent description using statistical ensembles on configuration space}

\author{M Reginatto$^{1}$ and M J W Hall$^{2}$}

\address{$^{1}$Physikalisch-Technische Bundesanstalt, Bundesallee 100, 38116 Braunschweig, Germany $^{2}$Theoretical Physics, Research School of Physics and Engineering, Australian National University, Canberra ACT 0200, Australia }

\ead{marcel.reginatto@ptb.de}

\begin{abstract}
We provide an overview of a canonical formalism that describes mixed quantum-classical systems in terms of statistical ensembles on configuration space, and discuss applications to measurement theory. It is shown that the formalism allows a general description of the measurement of a quantum system by a classical apparatus without running into inconsistencies. An example of classical and quantum particles interacting gravitationally is also given.
\end{abstract}

\section{Introduction}

There are a number of good reasons for modeling the interaction of quantum and classical systems. Some of them are practical: it is sometimes convenient for physical or computational reasons to model part of a physical system classically, in which case it is of interest to have a formalism that is free of inconsistencies and allows for well defined approximation schemes. But there are other reasons which stem from open problems in the foundations of physics. A good example can be found in measurement theory: in the standard Copenhagen interpretation of quantum mechanics, the measuring apparatus must be described in classical terms \cite{B1958,H1958} and this implies a coupling of some sort between a quantum system and the apparatus which is treated as a classical system. A measurement performed according to these principles will be discussed in section \ref{measurement}. Gravity provides another example: since there is no full quantum theory yet, one would like to know to what extent the quantization of gravity is forced upon us by consistency arguments alone \cite{AKR2008}, and one way to do this is to investigate how far we can get with a mixed system in which the gravitational field remains classical while matter is assumed to consist of quantum fields. This is a difficult problem that is beyond the scope of the paper. We will consider here a much simpler problem which may however be relevant to this issue, that of measurements where quantum and classical non-relativistic particles interact gravitationally; this is described in section \ref{gravity}.

There are, of course, a number of obstacles that have to be dealt with when modeling the interaction of quantum and classical systems. First of all, quantum mechanics and classical mechanics are formulated using very different mathematical structures, and it is necessary to to find a common mathematical framework before progress can be made. Furthermore, there are difficult conceptual issues, i.e. uncertainty principle, superposition, etc. What aspects of quantum and classical mechanics should be preserved in the description of a mixed system?

In the following sections we give an overview of a solution to this problem based on a canonical formalism that describes statistical ensembles on configuration space \cite{HR2005}. The formalism can be applied to both quantum and classical systems and allows a general and consistent description of interactions between them. This approach overcomes difficulties arising in previous attempts; in particular, the correct equations of motion for the quantum and classical sectors are recovered in the limit of no interaction, conservation of probability and energy are satisfied, uncertainty relations hold for conjugate quantum variables, and the formalism allows a back reaction of the quantum system on the classical system. A discussion of other approaches falls outside the scope of this paper, but see ref. \cite{HR2005} and the references quoted therein, and ref. \cite{H2008}.

\section{Ensembles in configuration space: the basic idea}

We start from the assumption that, as seems to be implied by quantum mechanics, the configuration of a physical system is an inherently statistical concept. The system is then described by an ensemble of configurations with probability density $P$, where $P \geq 0$ and $\int dx \,P(x,t) = 1 $. To derive equations of motion we introduce an {\it ensemble Hamiltonian} $\tilde{H}[P,S]$ where $S$ is canonically conjugate to $P$. The equations of motion take the form
\begin{equation}
\frac{\partial P}{\partial t} = \left\{ P,\tilde{H} \right\}_{PB}
= \frac{\delta\tilde{H}}{\delta S},~~~~~~~~~\frac{\partial S}{\partial t} = \left\{ S,\tilde{H} \right\}_{PB}
=-\frac{\delta\tilde{H}}{\delta P},
\end{equation}
where $\{ A,B \}_{PB}$ is the Poisson bracket of the fields $A$ and $B$.

The following ensemble Hamiltonians are of interest in that they lead to equations that describe the evolution of quantum and classical non-relativistic systems:
\begin{eqnarray}\label{HCandQ}
\tilde{H}_C[P,S] &=& \int dx\, P \left[ \frac{|\nabla S|^2}{2m} + V(x)\right] ,\\ \nonumber
{~} \\
\tilde{H}_Q[P,S] &=& \tilde{H}_C[P,S]
+  \frac{\hbar^2}{4} \int dx\ P\frac{|\nabla \log P|^2}{2m} . \nonumber
\end{eqnarray}
For example, the equations of motion derived from $\tilde{H}_Q[P,S]$ are given by
\begin{equation}\label{QEqMotion}
\frac{\partial P}{\partial t} + \nabla .\left( P\frac{\nabla S}{m} \right) =0,~~~\frac{\partial S}{\partial t} + \frac{|\nabla S|^2}{2m} + V +  \frac{\hbar^2}{2m}\frac{\nabla^2 P^{1/2}}{P^{1/2}} = 0
\end{equation}
while the equations of motion derived from $\tilde{H}_C[P,S]$ are the same as Eq. (\ref{QEqMotion}) but with $\hbar=0$. The first equation in Eq. (\ref{QEqMotion}) is a continuity equation, the second equation is the classical Hamilton-Jacobi equation when $\hbar = 0$ and a modified Hamilton-Jacobi equation when $\hbar \neq 0$. Defining $\psi:=\sqrt{P}~e^{iS/\hbar}$, Eq. (\ref{QEqMotion})  takes the form
\begin{equation}\nonumber
i\hbar \frac{\partial \psi}{\partial t}
= \frac{-\hbar^2}{2m}\nabla^2\psi + V\psi,
\end{equation}
which is the usual form of the Schr\"{o}dinger equation. Therefore, in this formalism, quantum and classical particles are treated on an equal footing, with differences being primarily due to the different forms of the respective ensemble Hamiltonians. Hence we have a framework that enables us to extend the formalism to mixed quantum-classical systems in a natural way, as discussed in the next section.

We end this section with some remarks concerning the interpretation of the canonically conjugate variables. While the interpretation of $P$ is straightforward, there are some subtle issues concerning the physical interpretation of $S$. To maintain full generality, $S$ should not be
regarded as a ``momentum potential''. In particular, for
an ensemble of classical particles with uncertainty described by
the probability $P$, it will not be assumed that the momentum
of a member of the ensemble is a well-defined quantity
proportional to the gradient of $S$, as it is done in
the usual deterministic interpretation of the Hamilton--Jacobi
equation. This avoids forcing a similar deterministic
interpretation in the quantum and quantum-classical cases. A
deterministic picture can be recovered for classical ensembles
precisely in those cases in which trajectories are operationally
defined \cite{HR2005}.

It is however possible to define local energy and momentum densities in terms of $S$. If $\tilde{H}[\lambda P, S] = \lambda \tilde{H}[P,S]$ (which is the case for the ensemble Hamiltonians that we consider in this article), then
\begin{equation} \nonumber
\tilde{H} = \int dx \,P\frac{\delta\tilde{H}}{\delta P} = - \int
dx\, P\frac{\partial S}{\partial t} = - \langle \partial S/\partial
t \rangle ,
\end{equation}
which shows that $\partial S/\partial t$ is a local energy density. Furthermore, $\int dx\, P\nabla S$ is the canonical infinitesimal generator of translations, since
\begin{eqnarray}
\delta P(x) = \delta \textbf{x} \cdot \left \{ P, \int dx\,
P\nabla S \right \}_{PB} = - \delta \textbf{x} \cdot \nabla P ,
\\ \nonumber
\delta S(x) = \delta \textbf{x} \cdot \left \{ S, \int dx\,
P\nabla S \right \}_{PB} = - \delta \textbf{x} \cdot \nabla S ,
\\\nonumber
\end{eqnarray}
under action of the generator, and therefore $P\nabla S$ can be considered a local momentum density. These results are generally valid; i.e. they hold true for classical, quantum and mixed systems.

\section{Interaction of quantum and classical systems: a configuration space description}

The formalism of configuration-space ensembles allows a general and consistent description of interactions between quantum and classical ensembles. We consider non-relativistic systems as an example here. A mixed quantum-classical ensemble Hamiltonian on a configuration space with coordinates $q$, $x$ is given by \cite{HR2005}
\begin{eqnarray}\label{HQC}
\tilde{H}_{QC}[P,S] &=& \int dq\,dx\, P\,\left[ \frac{|\nabla_x S|^2}{2M}
+ \frac{|\nabla_q S|^2}{2m}
 \right] \\
&+& \int dq\,dx\, P\,\left[  \frac{\hbar^2}{4} \frac{|\nabla_q \log P|^2}{2m} + V(q,x,t)\right]
\nonumber.
\end{eqnarray}
Here $q$ denotes the configuration space coordinate of a quantum particle of mass $m$ and $x$ that of a classical particle of mass $M$, and $V(q,x,t)$ is a potential energy function describing the quantum-classical interaction. The equations of motion for $P$ and $S$ derived from $\tilde{H}_{QC}$ are
\begin{eqnarray}\label{EqsCQ}
\frac{\partial P}{\partial t} &=&  -\nabla_q .\left( P
\frac{\nabla_qS}{m} \right) - \nabla_x.\left(P\frac{\nabla_xS}{M}\right), \\ \nonumber
{~} \\ \nonumber
\frac{\partial S}{\partial t} &=& -
\frac{|\nabla_qS|^2}{2m} - \frac{|\nabla_xS|^2}{2M} - V +
\frac{\hbar^2}{2m}\frac{\nabla_q^2 P^{1/2}}{P^{1/2}} .\nonumber
\end{eqnarray}

The approach based on configuration space ensembles overcomes difficulties arising in previous approaches. We collect here a number of properties which are proven in references \cite{HR2005,H2008}:

1. Probability and energy are conserved.

2. There is backreaction of the quantum system on the classical system.

3. The correct quantum and classical equations of motion are recovered in the limit of no interaction.

4. The equations satisfy Galilean invariance for potentials of the form $V(x,q,t)=V(|x-q|)$ in Eq. (\ref{EqsCQ}).

5. The two ``minimal requirements for a quantum-classical formulation'' specified by Salcedo \cite{CS1999,S2007} are satisfied: a Lie bracket may be defined on the set of observables, and this Lie bracket is equivalent to a Poisson bracket for classical observables and to $(i\hbar)^{-1}$ times the quantum commutator for quantum observables.

6. ``Configuration separability'' is satisfied: the classical configuration is invariant under any canonical transformation applied to the quantum component, and vice versa.

7. The ``definite benchmark ... for an acceptable quantum-classical hybrid system'' specified by Peres and Terno \cite{PT2001} is satisfied: the expectation values for the position and momentum observables of linearly-coupled quantum and classical oscillators obey the classical equations of motion.

8. Generalized Ehrenfest relations are satisfied.

We are not aware of any other formulation that satisfies all of these properties. In particular, requirements 5 is highly non-trivial \cite{CS1999,S2007,ACC2007}; it therefore provides a critical test for any formulation of mixed quantum-classical systems.

We end this section with some observations that concern the relation between the theory of ensembles in configuration space and the more familiar formulations of statistical theories. States are usually described by density operators in the case of quantum mechanics and by phase space distributions in the case of classical mechanics (the important case of a classical point particle corresponds to a special phase space distribution, i.e. a delta function, and hence it can be handled with the same formalism that is used for classical statistical theories). Fig. \ref{figure} shows the relation between phase space (classical mechanics), configuration space (classical, quantum mechanics) and the space of density operators (quantum mechanics). One of the reasons why the formalism of ensembles in configuration space is able to provide a consistent description of mixed quantum-classical systems has to do with the way in which these spaces are related. What is important here is that classical ensembles in configuration space are not as general as ensembles in phase space: in general, ensembles in phase space have to be described by \emph{mixtures} of classical ensembles in configuration space. This is analogous to the situation in quantum mechanics, where density matrices describe mixtures of pure states (wave functions). One therefore has the following relations:

\begin{figure}
\vspace{9pc}
\centerline{
\includegraphics[width=35pc]{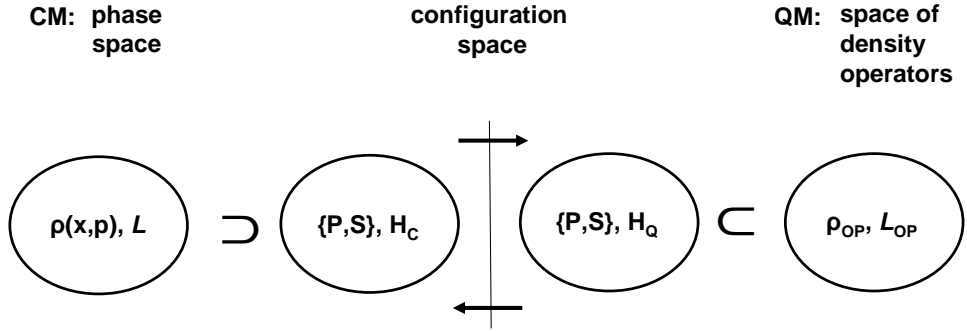}\hspace{2pc}
}
\caption{\label{figure}Relationship between phase space (classical mechanics), configuration space (classical, quantum mechanics) and the space of density operators (quantum mechanics).}
\end{figure}

a. Point particles $~\subset~$ Configuration space ensembles $~\subset~$ Phase space ensembles (CM).

b. Configuration space ensembles $~\subset~$ Density operators (QM).

c. Mixtures of configuration space ensembles $~=~$ Phase space ensembles (CM).

d. Mixtures of configuration space ensembles $~=~$ Density operators (QM).

In our formulation, mixed quantum-classical states are described in a unified way in terms of ensembles in configuration space, ensuring mathematical compatibility of the quantum and classical sectors. Furthermore, in this unified approach interactions between quantum and classical systems can be easily incorporated. By contrast, in other formulations attempts are made to bring together spaces that are mathematically incompatible, e.g. the quantum Hilbert space of wave functions is joined to either the space of point particles or the space of distributions in phase space of classical mechanics, and this leads to unsurmountable difficulties.

\section{Measurements of position}\label{measurement}

As an illustration of the application of the formalism to measurement theory, we consider a measurement of the position of a quantum particle carried out by a classical apparatus.

To interpret the outcome it will be helpful to introduce some additional concepts, which we now define. Consider the conditional probability $P(q|x) = P(q,x)/P(x) $. The conditional wave function is defined by
\begin{equation}
\psi(q|x) := \sqrt{P(q|x)}\, e^{i S(q,x)/\hbar}
,~~~~~~~|\psi_x\rangle := \int dq\,\psi(q|x)\,|q\rangle .
\end{equation}
We can also introduce a conditional density operator,
\begin{equation}
\rho_{Q|C} := \int dx\, P(x)\,|\psi_x\rangle\langle \psi_x| .
\end{equation}
It is important to keep in mind that $\psi(q|x)$ and $\rho_{Q|C}$ do \textit{not} satisfy linear Schr\"{o}dinger and Liouville equations, nor unitary invariance properties. These quantities only contain \textit{partial} information.

To model a measurement of position \cite{HR2005}, introduce the ensemble Hamiltonian
\begin{equation}
\tilde{H}_{\rm position} = \tilde{H}_{QC} + \kappa(t) \int dq\,dx\, P\, q.\nabla_x S .
\end{equation}
For an interaction over a short time period $[0,T]$ such that $\tilde{H}_{QC}$ can be ignored during the interaction,
\begin{equation}
\frac{\partial P}{\partial t} = -\kappa(t)\, q.\nabla_x P ,
~~~~~~ \frac{\partial S}{\partial t} = -\kappa(t)\, q.\nabla_x S ,
\end{equation}
which integrates to ($K=\int_0^T dt\,\kappa(t)$)
\begin{equation} \nonumber
P(q,x,T) = P(q, x-Kq, 0),~~~~~~S(q,x,T) = S(q, x-Kq,0) .
\end{equation}
Consider the case where the initial position of the pointer is sharply defined, $P(q,x,0) = \delta(x-x_0)P_Q(q)$. Then, after the measurement, $\Delta x \neq 0$, with probability $P_Q(q)$ of finding $x=x_0-Kq$. In other words, uncertainty is transferred from the quantum ensemble to the classical ensemble. The conditional density for the quantum component is diagonal in the position basis,
\begin{equation}
\rho_{Q|C} = \int dq \,P_Q(q)\, |q\rangle\langle q|
\end{equation}
and thus ``decoheres'' with respect to position.

This simple example shows that it is possible to describe measurements via quantum-classical interactions, as required for example by the Copenhagen interpretation, without running into inconsistencies. The main features are: (i) the measuring apparatus is described classically, as is required for the unambiguous communication and comparison of physical results; (ii) information about quantum ensembles is obtained via an appropriate interaction with an ensemble of classical measuring apparatuses, which correlates the classical configuration with a corresponding quantum property, and (iii) there is a conditional decoherence of the quantum ensemble relative to the classical ensemble, which depends upon the nature of the quantum-classical interaction.

For a more detailed analysis of this example, as well as an example of a measurement of spin, see ref.  \cite{HR2005}.

\section{Quantum and classical particles interacting gravitationally}\label{gravity}

In this section, we consider the scattering of two non-relativistic particles,
one a classical particle of mass $M$ (the
projectile) and the other one a quantum particle of mass $m$ (the
target). The interaction is assumed to be caused by the gravitational
attraction between the two particles. This scattering experiment constitutes
a measurement of the quantum system by a classical apparatus, in the sense that
information about the state of the quantum particle can be inferred from
measurements of the position of the classical particle made after the
interaction has taken place.

An interesting case is the one in which the quantum system is prepared
in such a way that the initial
amplitude for the quantum particle (i.e. as $t \rightarrow -\infty$ when the two particles are very far from each other so that the
interaction term can be neglected) has two peaks of equal magnitude, A
and B, that are well separated. Assume further that the classical particle is launched in the direction of peak A. What happens when the classical particle scatters from the quantum
particle? A ``naive approach'' suggests three possible mutually
exclusive outcomes: (a) the quantum particle is found at the location of
peak A and the classical particle comes very close to the quantum
particle of mass $m$: the scattering is very strong; (b) the quantum
particle is found at the location of peak B and the classical particle never
comes very close to the quantum particle of mass $m$: the scattering
is very weak; (c) as in semiclassical gravity \cite{PG1981}, the classical
particle ``sees'' a mass $m/2$ at the location of peak A and it
``sees'' a mass $m/2$ at the location of peak B: the scattering is
about one half of what one would calculate under assumption (a).
Each of these possibilities seems unrealistic.

The source of
difficulties is, clearly, the use of the ``naive approach'': it is
impossible to derive any reasonable conclusion without introducing a
concrete model for the mixed quantum-classical system, and we now proceed to do this.

The ensemble Hamiltonian will be of the form given in Eq. (\ref{HQC}) with interaction term
\begin{equation} \label{cqvg}
V(q,x) = -G\frac{mM}{|x-q|}.\nonumber
\end{equation}
The ensemble Hamiltonian that describes the system is then
\begin{eqnarray}
\tilde{H}_{QC}[P,S] &=& \int dq\,dx\, P\,\left[ \frac{|\nabla_q
S|^2}{2m} + \frac{|\nabla_x S|^2}{2M} \right.  \\
 &~& ~~~~~~~~~~~~ - \left. G\frac{mM}{|x-q|} +
\frac{\hbar^2}{4} \frac{|\nabla_q \log P|^2}{2m} \right]. \nonumber
\end{eqnarray}
It is convenient to introduce center-of-mass and relative coordinates $\overline{x}$ and $r$, and the total and relative masses $M_T$ and $\mu$. Then
\begin{eqnarray}
\tilde{H}_{QC} &=& \underbrace{ \int d\overline{x}dr \, P \left[
\frac{ |\nabla_{\overline{x}} S|^2}{2M_T} + \frac{\hbar^2 m}{4(m+M)}
\frac{|\nabla_{\overline{x}} \log P|^2}{2M_T} \right]}_{\textmd{(i)
quantum-like term:
free center-of-mass motion}} \\
&~& \nonumber\\
&+&  \underbrace{ \int d\overline{x}dr \, P \left[ \frac{ |\nabla_r
S|^2}{2\mu} + \frac{\hbar^2M}{4(m+M)} \frac{|\nabla_r \log
P|^2}{2\mu} -G\frac{\mu M_T}{|r|} \right] }_{\textmd{ (ii)
quantum-like term:
relative-motion in a potential}}\nonumber\\
&~& \nonumber\\
&-&  \underbrace{ \frac{2 \mu}{M_T} \int d\overline{x}dr \, P
\left[ \frac{\hbar^2M}{4(m+M)} \frac{\left( \nabla_{\overline{x}}
\log P \cdot \nabla_r \log P \right)}{2\mu} \right] }_{\textmd{(iii)
interaction term}}. \nonumber
\end{eqnarray}

The qualitative features of the solution can be determined without carrying out a
detailed calculation. To interpret this expression, compare to the ensemble Hamiltonian of a purely
quantum system; i.e. Eq. (\ref{HCandQ}), second line. We have a sum of three terms: (i) a quantum-like term corresponding
to free centre-of-mass motion but with a rescaled Planck constant
\begin{equation}
 \hbar_{\overline{X}} := [m/(m+M)]^{1/2}\, \hbar ;
\end{equation}
(ii) a quantum-like term corresponding to relative motion in
a potential $V(r)=-G\frac{\mu M_T}{|r|}$ but with a rescaled Planck constant
\begin{equation}
 \hbar_R := [M/(m+M)]^{1/2}\, \hbar ;
\end{equation}
and (iii) an intrinsic interaction term.

The predictions of the mixed quantum-classical system described here differ therefore
substantially from the outcomes predicted using the ``naive approach''. In particular,
one expects a solution with \emph{qualitative} features that resemble those of a purely
quantum system, however with important modifications induced by the
(iii) term and by the rescaling of the Planck constant in (i) and (ii). In particular, the classical and quantum configuration spaces will be correlated once the particles interact, as in the example of position measurement discussed previously. This will lead to a final state that is similar to an entangled state.
Cases (a) and (b) of the ``naive approach'' are ``either-or''
outcomes that are clearly not in agreement with this observation. Furthermore,
the theory that we have
used is fundamentally different from a theory built along the lines of standard semiclassical
gravity. Therefore, case (c) of the ``naive approach'' is also
excluded. A more extensive discussion of this system can be found in Appendix B
of ref. \cite{AKR2008}.

More generally, it is possible in principle to model the interaction of a classical gravitational field with
quantized matter fields \cite{HR2005,AKR2008,R2005}. This leads to a theory that is fundamentally different from semiclassical gravity.

\section{Concluding remarks}

It is seen that a consistent description of quantum-classical interactions is possible, based on a canonical formalism for ensembles in configuration space. Furthermore, this is the only description which satisfies all of properties 1-8 in section 3.

In the standard Copenhagen interpretation of quantum mechanics, it is assumed that the measuring apparatus is described classically and that information about a quantum system is obtained via an appropriate interaction between quantum system and classical apparatus. In this article, we have shown that the formalism of configuration-space ensembles allows a general and consistent description of the measurement of a quantum system by a classical apparatus. Using two examples, we have shown how this formalism is implemented ``in practice''.

The models described in this article provide illustrations of some qualitative features that one expects to be present in all such solutions: for example, uncertainty is transferred from the quantum system to the classical measuring apparatus, and there is decoherence of the quantum component. Concepts used to describe measurement in quantum theory will have their counterparts in this formalism, although perhaps with limited validity.

\section*{References}
\bibliography{Reginatto_Hall_DICE_2008}

\end{document}